\documentclass[sn-mathphys]{sn-jnl}

\jyear{2022}

\usepackage{amsmath,graphicx,hyperref,verbatimbox,array,float,doi}
\usepackage{indentfirst}
\usepackage{physics}
\usepackage{diagbox}
\usepackage{amssymb}
\usepackage{bbold}
\usepackage{colortbl}
\usepackage{multirow}
\usepackage{subcaption}
\newcommand{\figref}[1]{Fig.~\ref{#1}}
\newcommand{\tabref}[1]{Tab.~\ref{#1}}
\renewcommand{\eqref}[1]{Eq.~\ref{#1}}
\newcommand{\centered}[1]{\begin{tabular}{l} #1 \end{tabular}}

\begin{document}

\title[Quantum tomography of entangled qubits by time-resolved single-photon\dots]{Quantum tomography of entangled qubits by time-resolved single-photon counting with time-continuous measurements}

\author*[1]{\fnm{Artur} \sur{Czerwinski}}\email{aczerwin@umk.pl}

\affil*[1]{\orgdiv{Institute of Physics, Faculty of Physics, Astronomy and Informatics}, \orgname{Nicolaus Copernicus University in Torun}, \orgaddress{\street{ul. Grudziadzka 5}, \city{Torun}, \postcode{87-100}, \country{Poland}}}

\abstract{In this article, we introduce a framework for entanglement characterization by time-resolved single-photon counting with measurement operators defined in the time domain. For a quantum system with unitary dynamics, we generate time-continuous measurements by shifting from the Schr\"odinger picture to the Heisenberg representation. In particular, we discuss this approach in reference to photonic tomography. To make the measurement scheme realistic, we impose timing uncertainty on photon counts along with the Poisson noise. Then, the framework is tested numerically on quantum tomography of qubits. Next, we investigate the accuracy of the model for polarization-entangled photon pairs. Entanglement detection and precision of state reconstruction are quantified by figures of merit and presented on graphs versus the amount of time uncertainty.}


\maketitle

\section{Introduction}\label{sec1}

Quantum mechanics admits the existence of physical systems which demonstrate correlations that cannot be explained on the basis of classical physics \cite{Einstein1935}. In this context, we usually consider compound quantum systems which feature nonlocal quantum correlations that can be verified experimentally by detecting multiparticle quantum interference \cite{Bell1964}. In particular, quantum entanglement, which is a specific form of non-classical correlations, has been excessively studied with respect to creation, detection, and applications \cite{Horodecki1996,Horodecki2009}. Nowadays, we know that multipartite systems are not necessary to exhibit entanglement since one photon is sufficient to encode a Bell state \cite{Kim2003}

Entangled states are considered a key resource in quantum information theory \cite{Nielsen2000}. Particularly, entangled pairs of photons can be used for quantum key distribution (QKD) \cite{Ekert1991}. Other well-known applications of entangled states relate to: superdense coding \cite{Bennett1992,Mattle1996}, quantum teleportation \cite{Bennett1993,Anderson2020}, quantum computing \cite{Jozsa2003}, quantum interferometric optical lithography \cite{Boto2000}, etc. For these reasons, the ability to characterize entanglement based on measurements plays a crucial role in practical realizations of quantum protocols.

In the case of photons, quantum information can be encoded by exploiting different degrees of freedom, in particular: polarization, spectral, spatial, and temporal mode. Each approach requires distinct measurement schemes for state identification. In this work, we focus on two-photon polarization-entangled states.

State reconstruction of polarization-entangled photons, generated by a spontaneous-down-conversion photon source, was performed efficiently by polarization measurements \cite{White1999}. Then, photonic state tomography was developed in terms of both theory and experiments \cite{James2001,Altepeter2005}. Because each measurement is inherently inaccurate, we need to apply methods that produce reliable estimates of actual quantum states, such as maximum likelihood estimation (MLE) \cite{Hradil1997,Banaszek1999}. Different methods of quantum state estimation can be compared with respect to their accuracy \cite{Bantysh2021}.

For years, the problem of quantum state tomography (QST) has been intensively researched into, see, e.g., Ref.~\cite{dariano03,paris04,Rehacek2004,Quek2021}. Regardless of the particular QST method, we consider a total number of measurements as a resource. Therefore, economic frameworks, which aim at reducing the number of distinct measurement settings, are gaining in popularity. For example, dynamical maps can be utilized to generate an informationally complete set of measurement operators \cite{Czerwinski2021b}. There have also been proposals involving continuous measurement defined in the time domain, for example, by performing a weak (non-projective) continuous measurement on an ensemble of systems \cite{Silberfarb2005,Merkel2010,Smith2013}, including numerical optimization algorithms to reduce the influence of experimental noise \cite{Zhang2020}.

In this work, we consider time-continuous measurements generated by unitary dynamics from one operator. We gather data for state reconstruction by selecting a discrete set of time instants that correspond to moments of measurement. Since the precision of measurements is limited by the time resolution of the detector, we study the efficiency of the framework versus the amount of time uncertainty. Temporal uncertainty of the detector is considered a limiting factor in quantum state engineering. Therefore, theoretical models describing its formalism have been recently proposed \cite{Gouzien2018,Gouzien2019}.

In Sec.~\ref{measurements}, we define our measurements in the time domain. For a given unitary dynamics, we present possible trajectories which can be generated. Next, we discuss the impact of the detector jitter on the measurement operators. In Sec.~\ref{methods}, we introduce the QST framework with time-continuous measurements. Finally, in Sec.~\ref{results}, we present the results of numerical simulations for qubit reconstruction and entangled photon pairs tomography. In particular, for different numbers of photons per measurement, we investigate the efficiency of the framework versus the amount of time uncertainty.

\section{Time-continuous measurements}\label{measurements}

In the standard approach, one would need to perform a series of tomographically complete measurements to determine an unknown polarization state of a photon described by a $2 \times 2$ density matrix $\rho (0)$. An experimental setup would require: a polarizer, a quarter-wave plate (QWP), and a half-wave plate (HWP). The angles of the waveplates can be set arbitrarily, which allows the experimenter to measure the beam corresponding to selected polarizations. Traditionally, we utilize six polarization states: horizontal, vertical, diagonal, antidiagonal, right-circular, and left-circular, denoted by $\{\ket{H}, \ket{V}, \ket{D}, \ket{A}, \ket{R}, \ket{L}\}$. If the pair $\{\ket{V}, \ket{H}\}$ forms the standard basis, i.e.
\begin{equation}\label{2e0}
\ket{H} = \begin{pmatrix} 1 \\ 0  \end{pmatrix}, \hspace{1cm} \ket{V} = \begin{pmatrix} 0 \\ 1  \end{pmatrix},
\end{equation}
then the other states can be expressed as: $\ket{D} = (\ket{H} + \ket{V})/2$, $\ket{A} = (\ket{H} - \ket{V})/2$, $\ket{R} = (\ket{H} + i \ket{V})/2$, and $\ket{L} = (\ket{H} - i \ket{V})/2$. These six states can be divided into three pairs of orthogonal vectors, which correspond to mutually unbiased bases in the $2-$dimensional Hilbert space \cite{Wootters1989,Durt2010}. From these vectors, we can also generate a positive-operator valued measure (POVM), which can be considered an overcomplete measurement scheme.

In our framework, we assume that the photon undergoes a unitary evolution, which implies that:
\begin{equation}\label{2e1}
\rho (t) = U(t) \,\rho (0)\, U^{\dagger} (t),
\end{equation}
where $U(t)$ stands for a time-continuous unitary operator, i.e. $\forall \:t\geq0$ we have $U^{\dagger} (t) = U^{-1} (t)$ and the initial condition is satisfied: $U(0) = \mathbb{1}_2$.

A unitary evolution of a photonic state can be realized by utilizing the effects of stress-induced birefringence to create changes in the polarization of light traveling through a fiber. In particular, in-line polarization controllers create stress-induced birefringence within a single-mode fiber by mechanically compressing the fiber. This acts like a variable, rotatable waveplate. Both the angle and retardance of the waveplate can be continuously, independently adjusted, which allows any arbitrary input polarization state to be converted to any desired output polarization state. In other words, this method enables polarization control over the entire Bloch sphere. Therefore, by applying fiber polarization controllers, the polarization state entering the fiber can be transformed unitarily to any other polarization state on the Bloch sphere upon leaving the fiber. Subsequently, this implies the ability to implement an arbitrary continuous evolution of a polarization-encoded qubit by varying the amount of mechanical stress applied and rotation of the axis of birefringence.

Then, if initially we are able to perform one measurement characterized by an operator $M_{\xi} \geq 0$, we obtain, based on the Born rule, a time-dependent formula for the probability associated with this measurement:
\begin{equation}\label{2e2}
p_{\xi} (t) = \tr \left( M_{\xi} \,U(t) \,\rho (0)\, U^{\dagger} (t) \right) = \tr \left( U^{\dagger} (t) \,M_{\xi} \,U(t) \,\rho (0)  \right),
\end{equation}
where the latter expression is due to the cyclic property of the matrix trace. According to the Heisenberg representation, this implies that we can consider \eqref{2e2} as a measurement result for an evolving operator. Since $M_{\xi}$ is positive semi-definite, $\forall \:t\geq0$ we get $M_{\xi} (t) \equiv U^{\dagger} (t) \,M_{\xi} \,U(t) \geq 0$, which means that $M_{\xi} (t)$ fits to the concept of generalized measurements for any $t \geq 0$.

In general, an arbitrary $2 \times 2$ unitary matrix $U$ can be represented as \cite{Nielsen2000}:
\begin{equation}\label{2e3}
U = e^{i \alpha} \begin{pmatrix} e^{-i \frac{\beta}{2}} & 0 \\  \\ 0 &  e^{i \frac{\beta}{2}} \end{pmatrix} \begin{pmatrix} \cos \frac{\gamma}{2} & - \sin \frac{\gamma}{2}  \\  \\ \sin \frac{\gamma}{2} &  \cos \frac{\gamma}{2} \end{pmatrix}  \begin{pmatrix} e^{-i \frac{\delta}{2}} & 0 \\ \\ 0 &  e^{i \frac{\delta}{2}} \end{pmatrix},
\end{equation}
where $\alpha, \beta, \gamma$ and $\delta$ denote real parameters. The decomposition \eqref{2e3} provides an exact description of single-qubit operations by means of three rotation operators and a global phase shift. Since the global phase does not affect the probabilities, this part shall be omitted. In our model, we assume that the angles $\beta, \gamma$, and $\delta$ are linear functions of time, which allows us to define unitary evolution governed by an operator:
\begin{equation}\label{2e4}
U(t) := \begin{pmatrix} e^{-i \frac{\omega_{1} t}{2}} & 0 \\  \\ 0 &  e^{i \frac{\omega_{1} t}{2}} \end{pmatrix} \begin{pmatrix} \cos \frac{\omega_{2} t}{2} & - \sin \frac{\omega_{2} t}{2}  \\  \\ \sin \frac{\omega_{2} t}{2} &  \cos \frac{\omega_{2} t}{2} \end{pmatrix}  \begin{pmatrix} e^{-i \frac{\omega_{3} t}{2}} & 0 \\ \\ 0 &  e^{i \frac{\omega_{3} t}{2}} \end{pmatrix},
\end{equation}
where $\omega_{i} = 2 \pi/ T_i$.

\begin{figure}[h!]
		\centering
		\begin{tabular}{ c |c }
			\centered{$M_{\xi}$} &\centered{$M_{\xi} (t)$}\\\hline
			\centered{$\ket{H}\!\bra{H}$}&\begin{tabular}{l}\raisebox{- \height}{\includegraphics[width=0.3\columnwidth]{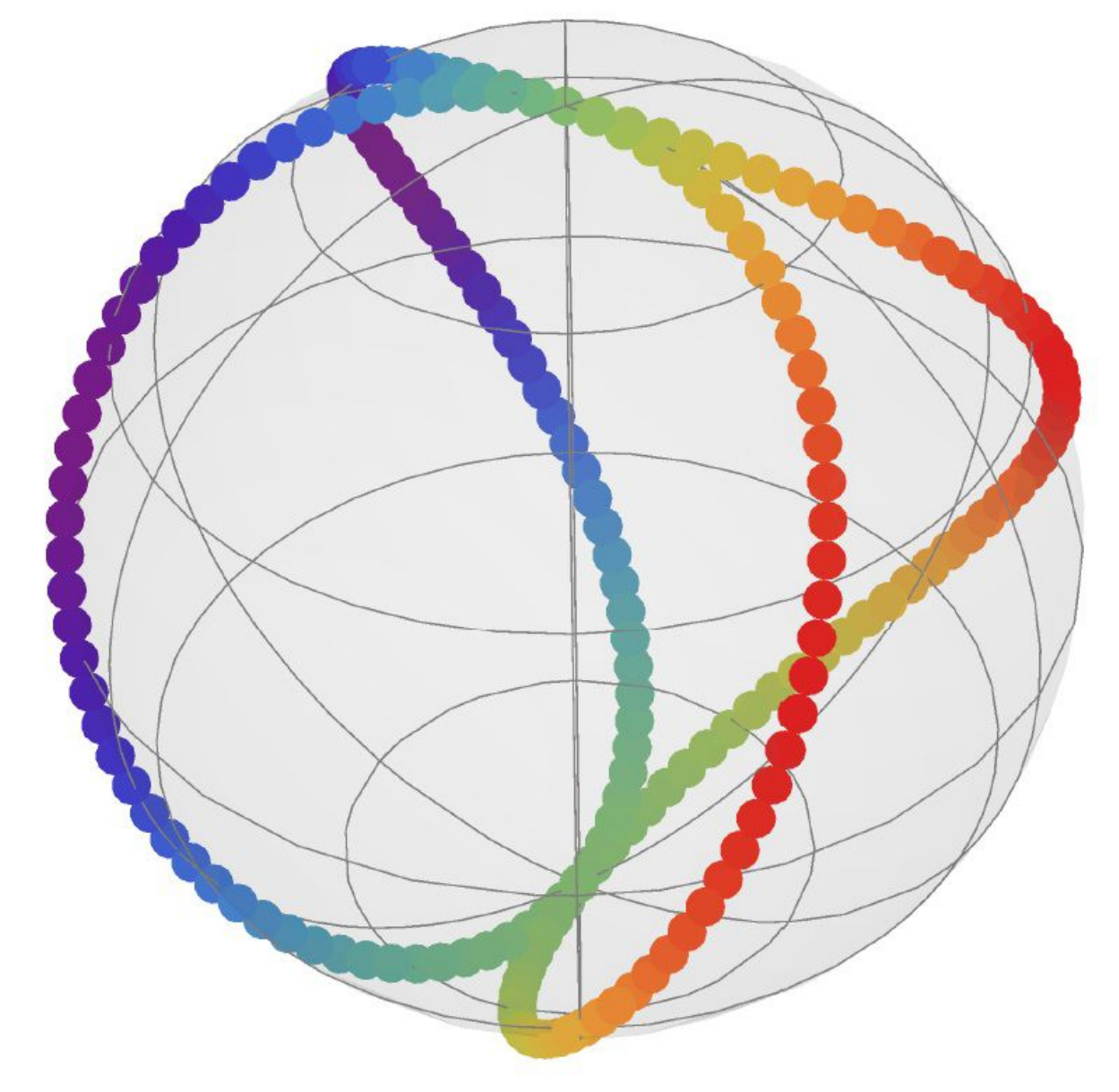}}\end{tabular} \\ \hline
			\centered{$\ket{D}\!\bra{D}$}&\begin{tabular}{l}\raisebox{-\height}{\includegraphics[width=0.3\columnwidth]{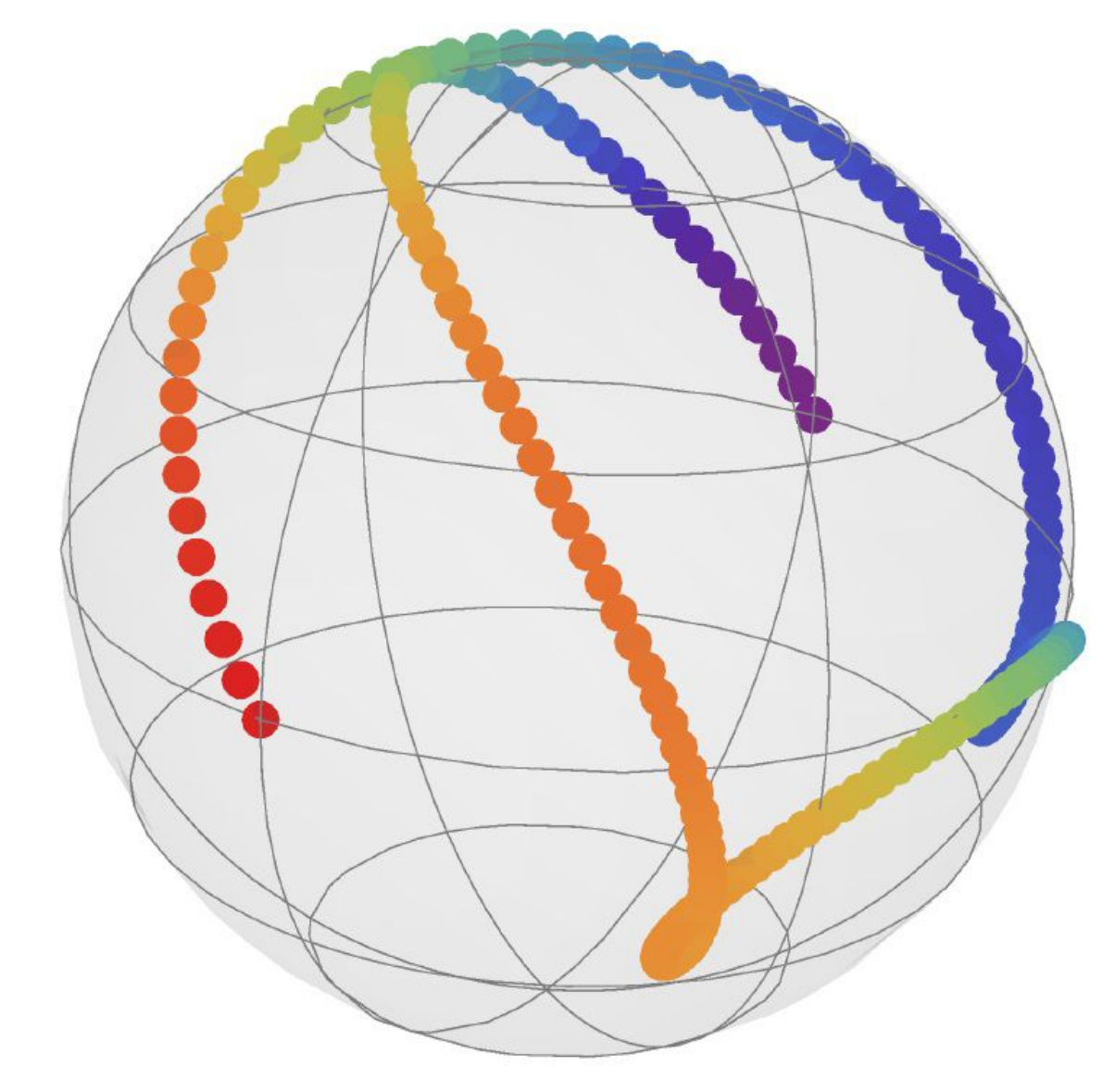}}\end{tabular}\\ \hline
			\centered{$\ket{R}\!\bra{R}$}&\begin{tabular}{l}\raisebox{-\height}{\includegraphics[width=0.3\columnwidth]{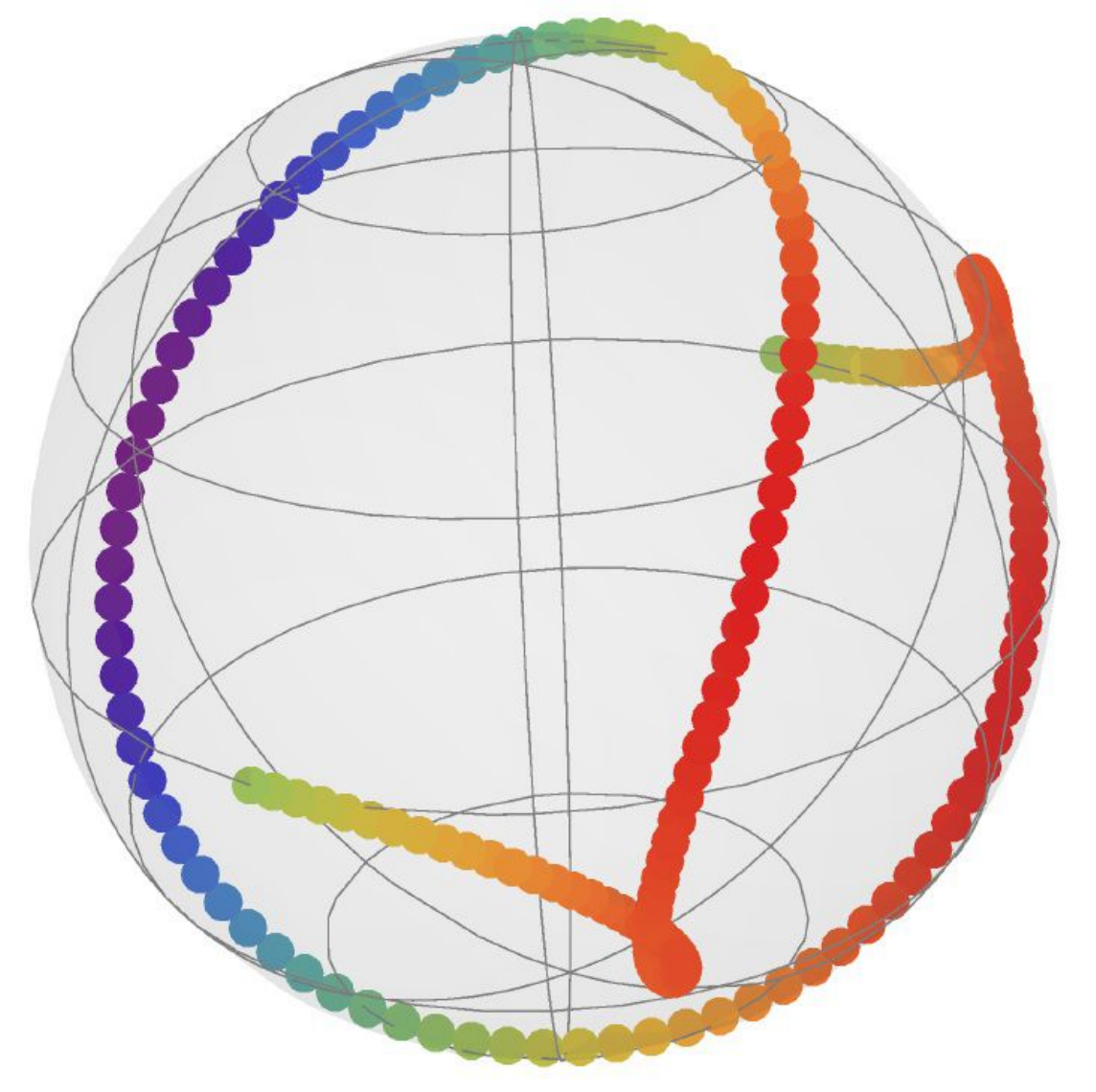}}\end{tabular}\\
		\end{tabular}
		\caption{Trajectories of time-continuous measurement operator $M_{\xi} (t)$ that can be generated from a given initial operator $M_{\xi}$ and a time-dependent unitary dynamics of the form \eqref{2e4}. The illustration was inspired by the presentation of POVMs in Ref.~\cite{SedziakKacprowicz2020}}
		\label{measurement1}
	\end{figure}

With \eqref{2e4} defining the time-evolution of a measurement operator $M_{\xi}$, we can follow the trajectory of $M_{\xi} (t)$ on the Bloch sphere for specific examples. Let us assume that $T_1 = 4 T,$ $T_2 = T$ and $T_3 = 2 T$, where $T$ denotes a period characterizing the dynamics. In \figref{measurement1}, one can observe the trajectories of $M_{\xi} (t)$ for three different initial operators $M_{\xi}$ in the time interval: $t \in [0, 2 T]$. 

In particular, if $M_{H} = \ket{H}\!\bra{H}$ we have:
\begin{equation}\label{2e5}
M_{H} (t) = \begin{pmatrix} \cos^2 \frac{\pi t}{T} & -\frac{1}{2} e^{i \pi t/T} \sin \frac{2 \pi t}{T} \\ \\ -\frac{1}{2} e^{-i \pi t/T} \sin \frac{2 \pi t}{T} & \sin^2 \frac{\pi t}{T} \end{pmatrix}.
\end{equation}
One can observe that $M_{H} (t)$ spans the Hilbert space. Furthermore, we can prove numerically that:
\begin{equation}\label{2e6}
\int_{0}^{2T} M_{H} (t) \, d t = \mathbb{1}_2,
\end{equation}
which is sufficient to conclude that $M_{H} (t)$ can be considered a time-continuous informationally complete POVM. From an experimental point of view, it implies that the apparatus should be in one setting corresponding to the horizontal polarization while the photon counting should be performed with respect to the arrival time of photons. In particular, if we select a discrete subset of six specific measurement operators, we can notice that:
\begin{equation}\label{2e7}
\begin{aligned}
\frac{1}{3} M_{H} (0)+ {}& \frac{1}{3} M_{H} (0.25 \,T)+ \frac{1}{3}M_{H} (0.5 \,T) + \frac{1}{3} M_{H} (0.75 \,T) +\\& +\frac{1}{3} M_{H} (1.25\, T)+ \frac{1}{3} M_{H} (1.75\, T)  = \mathbb{1}_2,
\end{aligned}
\end{equation}
which means that these six operators constitute an informationally complete POVM. The set of time instants selected in \eqref{2e7}, arranged in the increasing order, shall be denoted as $\mathcal{T} = \{t_1, t_2, \dots, t_6\}$. One can notice that the operators included in \eqref{2e7} are pairwise orthogonal, which additionally implies that they represent a scheme that is equivalent to the measurement of the six polarization states: $\{ \ket{H}, \ket{V}, \ket{D}, \ket{A}, \ket{R}, \ket{L} \}$.

	\begin{figure}[h!]
		\centering
\renewcommand{\arraystretch}{1.4}
		\begin{tabular}{c|c}
			\centered{$\sigma_j$} &\centered{$\widetilde{M}_{H} (t) $}\\\hline
			\centered{$0.1\,T$}&\begin{tabular}{l}\raisebox{-\height}{\includegraphics[width=0.25\columnwidth]{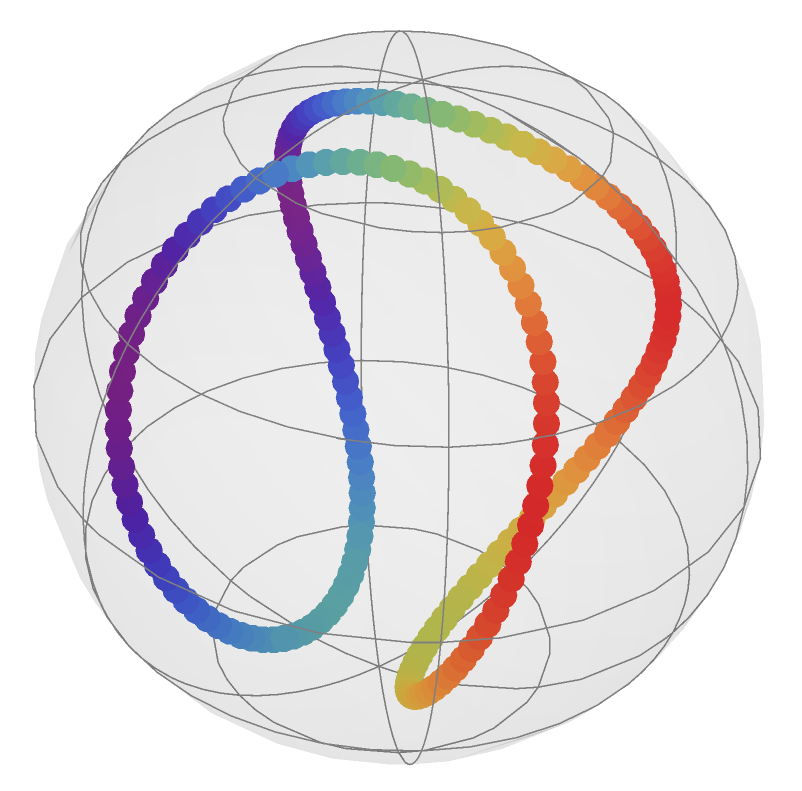}}\end{tabular} \\ \hline
			\centered{$0.2 \, T$}&\begin{tabular}{l}\raisebox{-\height}{\includegraphics[width=0.25\columnwidth]{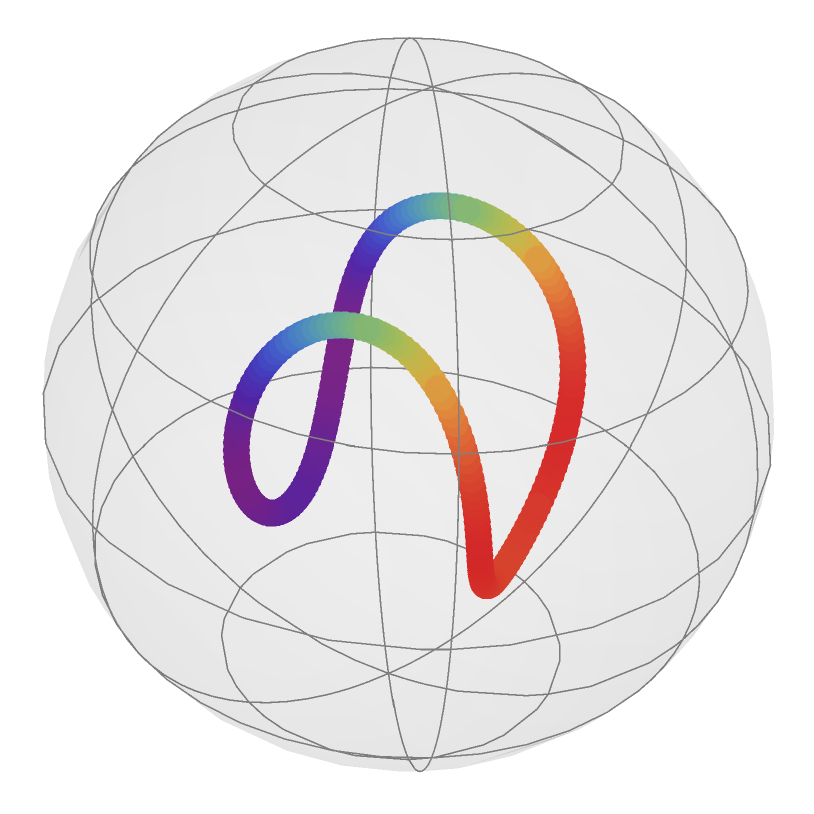}}\end{tabular}\\ \hline
			\centered{$0.3\, T$}&\begin{tabular}{l}\raisebox{-\height}{\includegraphics[width=0.25\columnwidth]{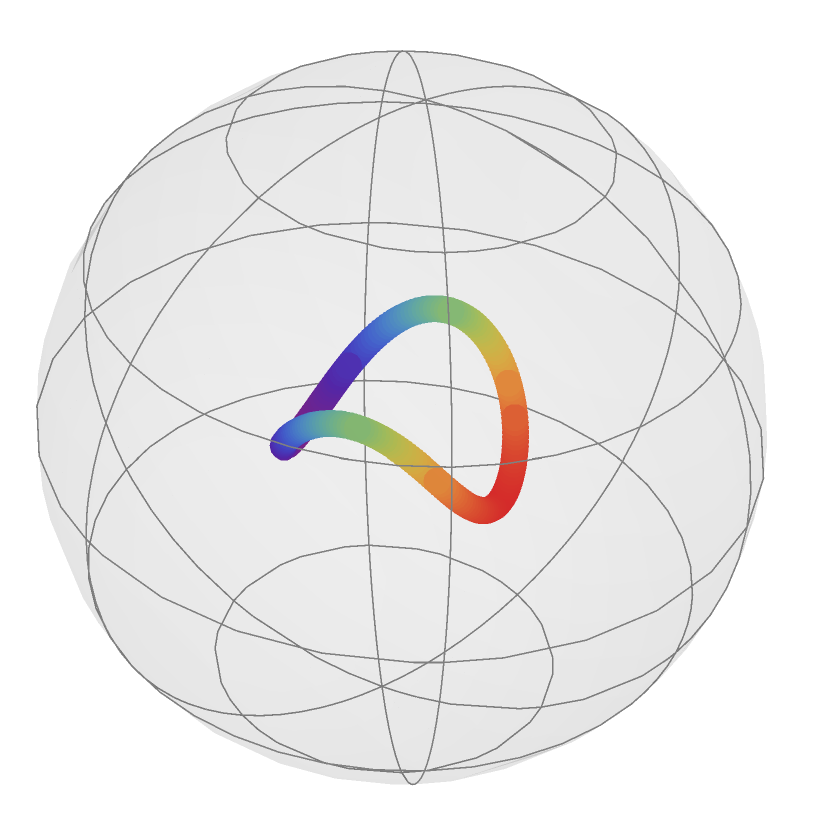}}\end{tabular}\\ \hline
			\centered{$0.5\, T$}&\begin{tabular}{l}\raisebox{-\height}{\includegraphics[width=0.25\columnwidth]{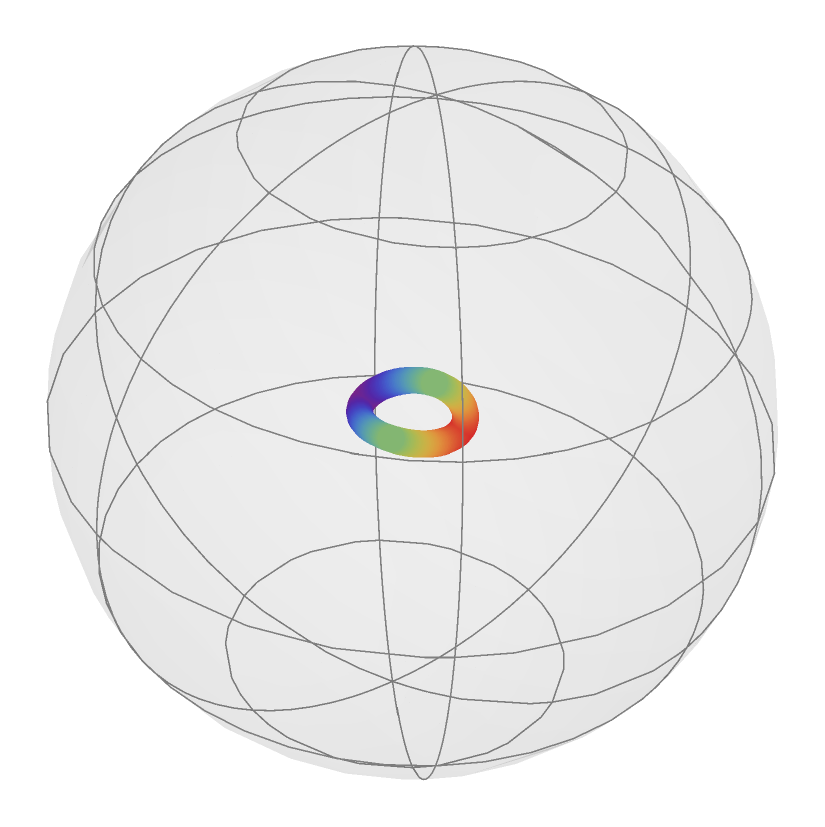}}\end{tabular}\\ \hline
			\centered{$0.75\, T$}&\begin{tabular}{l}\raisebox{-\height}{\includegraphics[width=0.25\columnwidth]{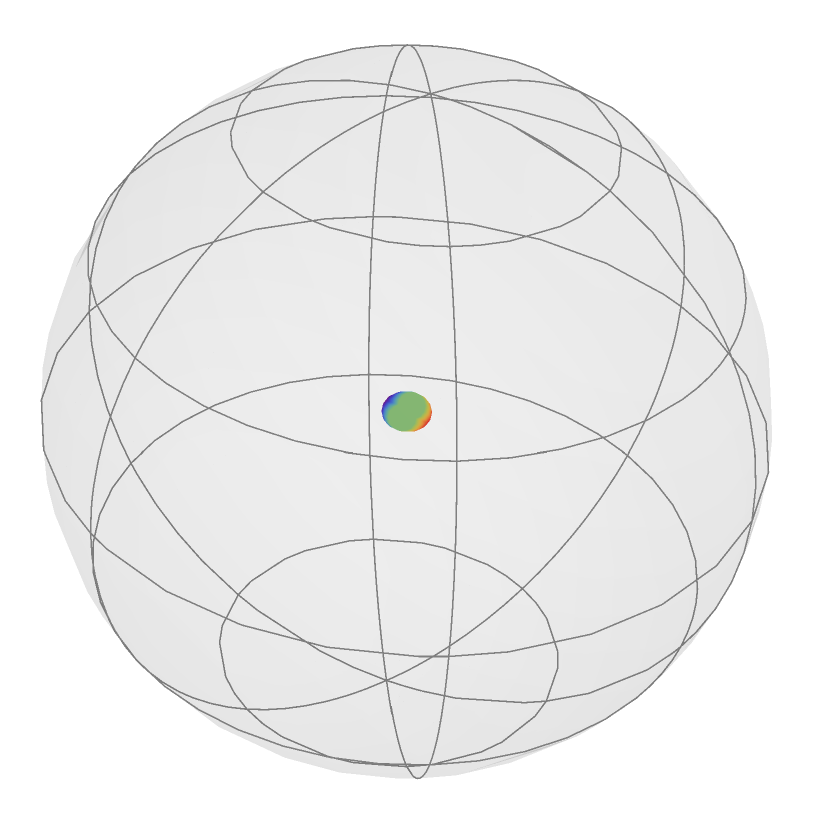}}\end{tabular}\\ 
		\end{tabular}
		\caption{Trajectories of realistic measurement operators $\widetilde{M}_{H} (t) $ for selected values of $\sigma_j$. The form was inspired by the presentation of POVMs in Ref.~\cite{SedziakKacprowicz2020}}
		\label{measurement2}
	\end{figure}

In a realistic scenario, we have to take into account errors connected with measurements. In the case of measurements that are performed in the time domain, one needs to consider the time uncertainty associated with the detector. Every detector features a timing jitter that has a detrimental impact on the time-resolved photon counting. This process can be modeled by a Gaussian distribution given as \cite{SedziakKacprowicz2020}:
\begin{equation}\label{2e55}
q_j (t) := \frac{\exp \left(- \frac{t}{2 \sigma_j ^2}  \right)}{\sqrt{ 2 \pi \sigma_j ^2}},
\end{equation}
where $\sigma_j$ stands for the timing jitter. Then, the measurement operator distorted by the timing uncertainty, denoted by $\widetilde{M}_{\xi} (t)$, can be expressed by a convolution \cite{SedziakKacprowicz2020}:
\begin{equation}\label{eqjitter}
\widetilde{M}_{\xi} (t) = \int_{- \infty}^{\infty} M_{\xi} (\tau)\, q_j (t- \tau) \,d \tau.
\end{equation}

The impact of the timing uncertainty depends on the value of $\sigma_j$. If we possess state-of-the-art technology, we can perform reliable and precise measurements. To demonstrate the distortion of the measurement operators due to the jitter, we consider $M_{H} (t)$ since this operator represents an informationally complete POVM. In \figref{measurement2}, one can observe how the quality of the measurement degenerates as we increase the value of $\sigma_j$. The trajectory of $\widetilde{M}_{H} (t)$ gets squeezed, and the purity of the operators decreases if we add more time uncertainty. For $\sigma_j = 0.75 \,T$, the trajectory resembles only a small ball located in the center. The figure shrinks to a single point if we further increase $\sigma_j$.

In addition, the temporal shape of photons is another quantity that could influence the process of photon counting in this scenario. The temporal probability density of a single photon can be modeled by a Gaussian distribution, cf. Ref.~\cite{SedziakKacprowicz2020}. Thus, the impact of this element on the measurement scheme could be considered analogously as in \eqref{2e55}-\ref{eqjitter}. However, in the present work, we neglect the consequences of the temporal width of photons because for a femtosecond laser, the ratio between the timing jitter and the standard deviation of the photon's temporal distribution could come up to $10^3$. Therefore, the detector's timing uncertainty is considered the key limiting factor that affects the framework.

\section{Framework for quantum state tomography}\label{methods}

\subsection{Methods: qubits}

In our QST framework, we assume that the experimental setup is adjusted so that we measure the photon count corresponding to the horizontal polarization, i.e., the apparatus performs the projective measurement defined by the operator $M_{H} = \ket{H}\!\bra{H}$. The setting is fixed, and we neglect any errors due to angular uncertainties. However, due to fiber polarization controllers prior to the measurement, the polarization state of photon changes unitarily according to \eqref{2e1}. Thus, the photon counting has to be performed in the time domain, which implies that we utilize a time-resolved detector for photons with different arrival times. Each act of measurement is performed for a beam consisting of, on average, $\mathcal{N}$ photons generated in the same polarization state. In other words, the source produces identical photons, but as they travel through a fiber, they undergo a unitary evolution according to \eqref{2e4}.

Based on the notation introduced in Sec.~\ref{measurements}, we can write a formula for the expected photon count:
\begin{equation}\label{3e1}
n_E (t) = \mathcal{N} \, \tr \left( M_{H} (t) \, \rho \right),
\end{equation}
where $\rho$ stands for an unknown density matrix that represents the quantum state of photons. Since we have no a priori knowledge about the state, we follow the Cholesky decomposition, which allows us to write \cite{James2001,Altepeter2005}:
\begin{equation}\label{3e2}
\rho = \frac{W^{\dagger} W}{\tr\: (W^{\dagger} W)}, \hspace{0.35cm}\text{where}\hspace{0.35cm}W=\begin{pmatrix}w_1 & 0 \\ w_3 + i\,w_4 & w_2 \end{pmatrix}.
\end{equation}
Cholesky factorization guarantees that the result of QST framework shall be physical, i.e. $\rho$ is positive semidefinite, Hermitian, of trace one. To reconstruct the density matrix $\rho$, one needs to determine the real parameters $\mathcal{W} \equiv \{w_1, w_2, w_3, w_4\}$.

On the other hand, the measured photon counts are computed by implementing the operator $\widetilde{M}_{H} (t) $ distorted by the time uncertainty. Additionally, since the framework is founded on photon-counting, we impose the Poisson noise \cite{Hasinoff2014}, which is typically taken into account in such a scenario, see e.g. Ref.~\cite{SedziakKacprowicz2020,Czerwinski2021}. Thus, we obtain a formula for the measured photon counts:
\begin{equation}\label{3e3}
n_M (t) = \mathcal{N}' \, \tr \left( \widetilde{M}_{H} (t) \, \rho_{in} \right),
\end{equation}
where $\mathcal{N}' $ for each act of measurement is generated randomly from the Poisson distribution characterized by the mean value $\mathcal{N}$, i.e. $\mathcal{N}' \in \mathrm{Pois} (\mathcal{N})$. The input states, $\rho_{in}$, are constructed from a general representation of the density matrix of a qubit:
\begin{equation}\label{3e4}
\rho_{in} = \frac{1}{2} \begin{pmatrix} 1 + r \,\cos \theta & r \:\Delta \\ \\ r \:\overline{\Delta} & 1 - r \,\cos \theta   \end{pmatrix},
\end{equation}
where $\Delta  = \sin \theta \cos \phi - i \sin \theta \sin \phi$ and $\overline{\Delta}$ denotes the complex conjugate of $\Delta$. In our framework, we are able to generate numerically experimental data for any input state. We shall consider a sample of input states such that the full range of all parameters is covered, i.e. $0\leq r \leq 1$, $0 \leq \phi < 2 \pi$ and $0\leq \theta \leq \pi$. Having a sample of input states is necessary to evaluate the average performance of the method.

Each input state goes through the tomographic scheme, and its density matrix is reconstructed by maximum likelihood estimation (MLE) \cite{Hradil1997,Banaszek1999}. We minimize the likelihood function of the form \cite{Ikuta2017}:
\begin{equation}\label{3e5}
\mathcal{L}(\mathcal{W}) = \sum_{k=1}^{6} \left[\frac{\left(n_M (t_k) -  n_E (t_k) \right)^2}{ n_E (t_k) }  + \ln n_E (t_k) \right],
\end{equation}
where we applied the discrete set of time instants $\mathcal{T}$, cf. \eqref{2e7}.

\subsection{Methods: entangled qubits}\label{methequbits}

We assume that our source can produce polarization-entangled photons in spontaneous parametric down-conversion (SPDC) process that converts one photon into a pair of photons \cite{Klyshko1970,Burnham1970}. Then, by $\mathcal{N}$ we denote the average number of photon pairs produced by the source. Each photon travels in one arm of the experimental setup and can be measured by the detector with a different arrival time. Thus, for an entangled photon pair, we apply two-qubit measurement operators that are the tensor products of single-qubit operators. We obtain:
\begin{equation}\label{3e6}
M^{2q}_{H} (t_i, t_j) := M_{H} (t_i) \otimes M_{H} (t_j),
\end{equation}
where $t_i, t_j \in \mathcal{T}$. Analogously, we define a two-qubit measurement operator burdened with time uncertainty, denoted by $\widetilde{M}^{2q}_{H} (t_i, t_j)$. Since we consider all combinations of time instants from $\mathcal{T}$, we get $36$ two-qubit measurement operators, which allows us to generate coincidence counts burdened with time uncertainty. An overcomplete set with $36$ measurement operators is typically used for practical implementations of QST protocols, see, e.g., Ref.~\cite{Horn2013}.

The formula for expected photon counts for entangled qubits is defined in the same vein as for qubits, cf. \eqref{3e1}. However, the lower triangular matrix $W$, which appears in the Cholesky decomposition, now comprises $16$ real parameters:
\begin{equation}\label{3e6}
W = \begin{pmatrix} w_1 & 0 & 0 &0 \\ w_5 + i\, w_6 &  w_2 & 0 &0 \\  w_{11} + i \,w_{12} & w_7 + i\, w_8 & w_3 &0 \\ w_{15} + i\, w_{16} & w_{13} + i\, w_{14} & w_9 + i\, w_{10} & w_4 \end{pmatrix}.
\end{equation}
On the other hand, measured photon counts are calculated analogously to \eqref{3e3} with $\rho_{in}$ standing for a maximally entangled Bell-type state, i.e. $\rho_{in} = \ket{ \Phi (\alpha)}\! \bra{\Phi (\alpha)}$, where:
\begin{equation}\label{3e7}
\ket{\Phi (\alpha)} = \frac{1}{\sqrt{2}} \left(\ket{00} + e^{i \alpha} \ket{11} \right),
\end{equation}
where $\{\ket{00}, \ket{01}, \ket{10}, \ket{11}\}$ denotes the standard basis in $4-$dimensional Hilbert space and $\alpha \in [0, 2 \pi)$ stands for the relative phase. In our model, we select a sample of input states of the form \eqref{3e7} with the relative phase covering the full range. Then, for each input state, we generate measured photon counts from $36$ measurement operators $\widetilde{M}^{2q}_{H} (t)$ with the Poisson noise. Finally, each state is reconstructed -- by minimizing the likelihood function, we obtain the values of the parameters $w_1, w_2, \dots, w_{16}$ that fit optimally to the simulated experimental data.

\subsection{Performance analysis}

The goal of the research is to quantify the efficiency of the measurement scheme introduced in Sec.~\ref{measurements}. First, we test it on qubits and then on entangled qubits. Each time, we select a sample of quantum states. For every input state $\rho_{in}$ we generate noisy measurements and perform tomographic reconstruction by MLE. Then, the original state is compared with its estimate, $\rho_{out}$, by computing the value of quantum fidelity \cite{Uhlmann1986,Jozsa1994,Bengtsson2006}:
\begin{equation}\label{3e8}
\mathcal{F} := \left(\tr \sqrt{\sqrt{\rho_{out}} \, \rho_{in}  \, \sqrt{\rho_{out}}} \right)^2.
\end{equation}
Finally, we calculate the average fidelity over the sample, denoted by $\mathcal{F}_{av}$, which is the figure of merit quantifying the accuracy of the measurement scheme, cf. Ref.~\cite{SedziakKacprowicz2020,Czerwinski2021,Czerwinski2021a}. In addition, the sample standard deviation (SD) is provided to quantify the amount of statistical dispersion.

To better characterize the efficiency of the framework in the case of qubits, we consider a sample of pure pairs of orthogonal states: $\{\rho_{in}, \rho_{in}^{\perp}\}$. Each state goes individually through the framework, and we obtain a pair of corresponding estimates: $\{\rho_{out}, \rho_{out}^{\perp}\}$. For the outcomes of the QST framework, we compute the trace distance \cite{Nielsen2000,Bengtsson2006}:
\begin{equation}\label{3e9}
D (\rho_{out}, \rho_{out}^{\perp}) = \frac{1}{2} \tr \|\rho_{out} - \rho_{out}^{\perp}\|,
\end{equation} 
where $\|X\| \equiv \sqrt{X X^{\dagger}}$. Since we consider a sample of pure orthogonal states, we know that $D (\rho_{in}, \rho_{in}^{\perp})=1$. Thus, the value of \eqref{3e9} quantifies the distinguishability of the quantum states delivered by the QST framework. Finally, for the sample of pairs, we calculate the mean trace distance, $D_{av}$, which indicates how well, on average, orthogonality is preserved by the measurement scheme.

For entangled states, we compute the concurrence, $C[\rho]$, which quantifies the amount of entanglement \cite{Hill1997,Wootters1998}. This figure is related to another entanglement measure -- entanglement of formation \cite{Bennett1996}. For any quantum state $\rho$ the concurrence satisfies: $C[\rho] \in [0, 1]$, where $C[\rho] =0$ for separable states and $C[\rho] = 1$ for maximally entangled states. The fact that the concurrence can be considered an entanglement monotone implies that it can be applied to quantify the amount of entanglement detected by a measurement scheme, see e.g., Ref.~\cite{Walborn2006,Buchleitner2007,Neves2007,Bergschneider2019}. In our framework, we consider a sample of maximally entangled states of the form \eqref{3e7}. Then, for each density matrix resulting from the tomographic scheme, $\rho_{out}$ we calculate the concurrence: $C[\rho_{out}]$. Next, the average concurrence is computed over the sample, denoted as $C_{av}$, in order to evaluate in general the performance of the framework in entanglement detection (each result is accompanied by the corresponding SD to measure the variance in the sample).

\section{Results and analysis}\label{results}

\subsection{Qubit tomography}

First, we select a sample of $8820$ qubits according to \eqref{3e4} with parameters $r, \theta$, and $\phi$ covering the whole Bloch ball. To each state from the sample, we apply the QST framework. Then, for the sample, we compute the average fidelity and the corresponding SD. In \tabref{qubitadata1}, in the columns with the heading "Mixed", one finds the results for different values of the jitter ($\sigma_j$) and numbers of photons ($\mathcal{N}$).

We observe three tendencies that were anticipated. First, if we analyze the columns, we notice that the average fidelity declines as we increase the jitter, whereas the corresponding SD grows. This feature is attributed to the fact that the measurement operators are squeezed toward the center of the Bloch ball for a greater time uncertainty, see \figref{measurement2}. Second, by comparing the figures in rows, we see that the results improve if we increase the number of photons per measurement. This was also anticipated since the number of photons is negatively correlated with the influence of the Poisson noise. Thirdly, we observe that the results obtained based on low-photon statistics exhibit a large variance, which implies that the fidelities are more scattered.

One may argue that the decline in the accuracy of state estimation is not so significant when compared with the collapse of the measurement operators, as presented in \figref{measurement2}. This comes from the fact that, with the squeezed measurement operators, we can still properly estimate the states which lie inside of the Bloch ball. However, in many applications, we do not implement states of a great entropy but utilize only pure states, e.g., to encode information or in QKD protocols. Therefore, it appears justifiable to measure the accuracy of the framework for a sample of pure states.

\begin{table}[h]
\centering
\setlength{\tabcolsep}{3.5pt} 
\renewcommand{\arraystretch}{2.45}
	\begin{tabular}{|c|c|c|c|c|c|c|}
\hline
\multirow{2}{*}{
				\backslashbox[9 mm]{$\sigma_j$}{$\mathcal{N}$}} & \multicolumn{2}{c|}{$10$} & \multicolumn{2}{c|}{$100$} & \multicolumn{2}{c|}{$1\,000$}\\ \cline{2-7} 
	 	&  Mixed & Pure & Mixed & Pure & Mixed & Pure \\ \hline

0 &$0.94(5)$ & $0.98(2)$  &$0.996(4)$ & $0.995(5)$ & $0.998(2)$ & $0.997(2)$   \\ \hline
0.1\,T &$0.94(6)$ & $0.91(8)$ &$0.98(2)$ & $0.91(3)$ & $0.98(2)$ & $0.91(2)$  \\ \hline
0.2\,T &$0.91(9)$ & $0.74(11)$ &$0.95(5)$ & $0.74(4)$ & $0.95(5)$ & $0.74(4)$  \\ \hline
0.3\,T &$0.88(12)$ & $0.62(13)$ &$0.92(8)$ & $0.62(5)$  & $0.92(8)$ & $0.62(5)$  \\ \hline
0.4\,T & $0.86(13)$ & $0.56(14)$ & $0.9(1)$ & $0.57(5)$ & $0.9(1)$ & $0.57(5)$ \\ \hline
0.5\,T & $0.85(14)$ & $0.54(15)$ & $0.89(11)$ & $0.54(5)$ & $0.89(11)$ & $0.54(3)$ \\ \hline
	\end{tabular}
	\caption{Average fidelity with SD in QST of qubits in different measurement scenarios. "Mixed" refers to a sample of $8820$ input states which cover the whole Bloch ball whereas "Pure" refers to a sample of $420$ states which lie on the Bloch sphere.}
	\label{qubitadata1}
\end{table}

In \tabref{qubitadata1}, in the columns with the heading "Pure", one finds the results for a sample $420$ pure states, i.e., we substitute $r=1$ to \eqref{3e4} and the other parameters ($\theta$ and $\phi$) take a set of values to cover the Bloch sphere. Apart from the same tendencies as already described, we can formulate two other conclusions. If we analyze $\sigma_j = 0$ (ideal measurements), then we see that the QST framework is more accurate in pure state reconstruction for the lowest number of photons (i.e., $\mathcal{N} =10$) than it is for mixed states. However, if we consider a non-zero value of the detector jitter, we notice that the quality of pure state estimation is substantially lower than in the case of a general sample of qubits. These results confirm the hypothesis that as the measurement operators shrink, they become less efficient at pure state estimation.

\begin{figure}[h]
	\centering
	\begin{tabular}{c}
		\centered{\includegraphics[width=0.75\columnwidth]{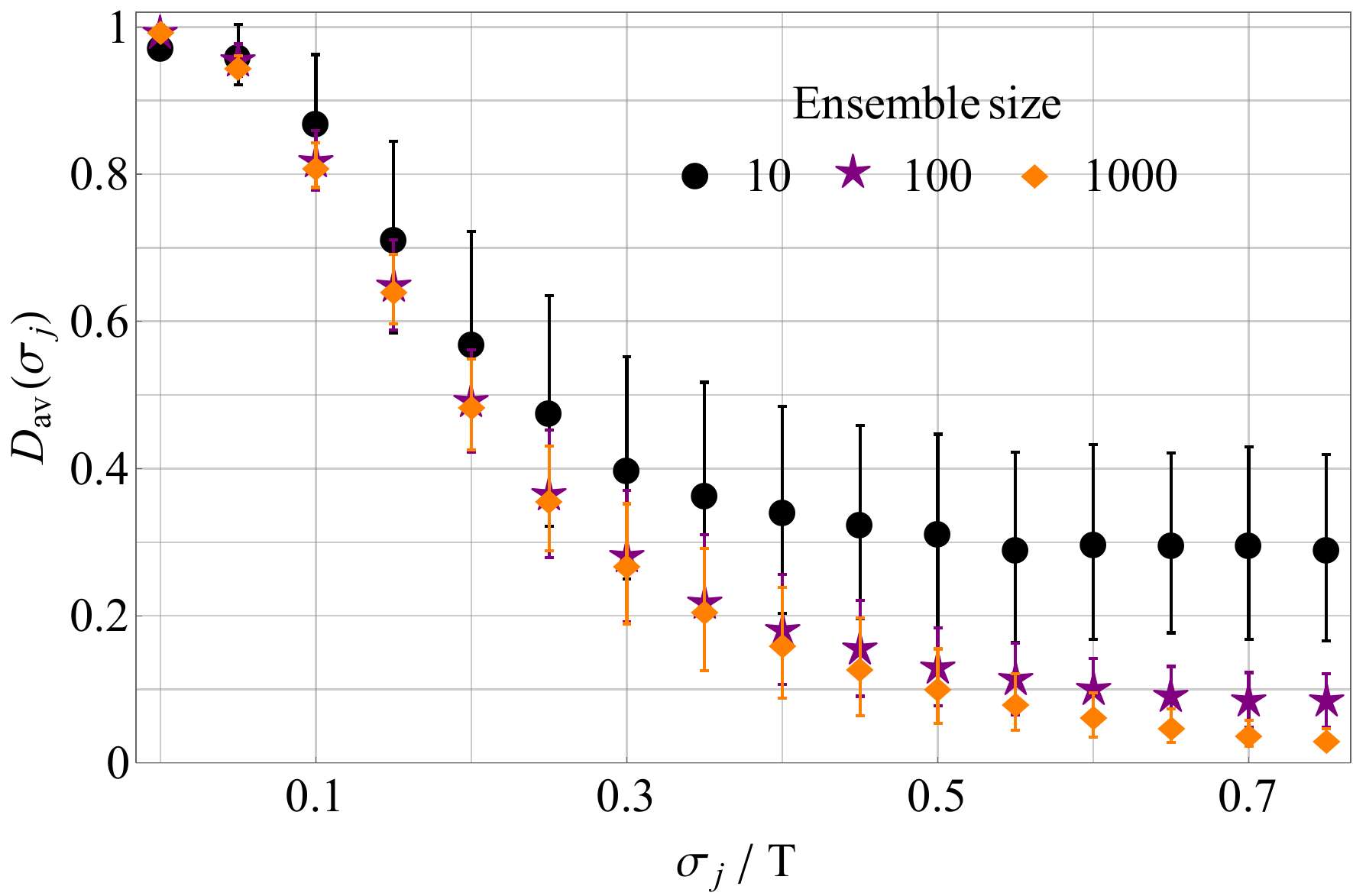}}
	\end{tabular}
	\caption{Plots of $D_{av} (\sigma_j)$ for a sample of $210$ pairs of orthogonal qubits. Three different numbers of photons per measurement were considered. Error bars correspond to one SD.}
	\label{qubitdata2}
\end{figure}

Next, the sample of $420$ pure states is divided into $210$ pairs of orthogonal states. For each pair of estimates resulting from the tomographic technique, we compute the trace distance. In \figref{qubitdata2}, one can observe the average trace distance for the sample, $D_{av} (\sigma_j)$, versus the amount of time uncertainty quantified by $\sigma_j$. The plots were generated for three different numbers of photons produced by the source per measurement (ensemble size). The most reliable plot corresponds to the greatest number of photons since, in such a case, the errors related to photon-counting are the least severe. For $\mathcal{N} = 1\,000$, we can observe how the average trace distance between the estimated states decreases as we add more time uncertainty into the measurements. In particular, if $\sigma_j=0$, we observe that the orthogonality is perfectly preserved. On the other hand, for $\sigma_j = 0.75 \,T$ the average trace distance is very close to zero. This means that the framework cannot differentiate between orthogonal input states for higher values of the detector jitter. These numerical results are in agreement with \figref{measurement2}, where it was shown that for $\sigma_j = 0.75 \,T$ the trajectory of $\widetilde{M}_{H} (t) $ shrinks into a point in the center of the Bloch sphere. Thus, such a measurement scheme, produces estimates which lie in the region of the maximally mixed state and, for this reason, the distance between the estimates of orthogonal states is close to zero.

Finally, a separate remark should be made concerning the performance of the measurements when $\mathcal{N}=10$. We observe that the plot for this case lies above the other two. One would think that this implies that the scenario with the lowest number of photons has an advantage over the other approaches when it comes to the distinguishability between quantum states. However, this effect is deceptive. There is no rational reason to assume that if the measurement operators are shrunk into a point, we can perform better by utilizing fewer photons. In such a case, the plot for $\mathcal{N}=10$ should converge to zero as the other two. In addition, we observe that the results corresponding to $\mathcal{N}=10$ feature a great deal of variance, which was indicated in \figref{measurement2} by error bars representing one SD. Therefore, this effect should be attributed to the Poisson noise, which distorts the measured counts significantly and leads to dysfunctional state estimation.

\subsection{Entangled qubits tomography}

\begin{figure}[h]
	\centering
		\centered{\includegraphics[width=0.75\columnwidth]{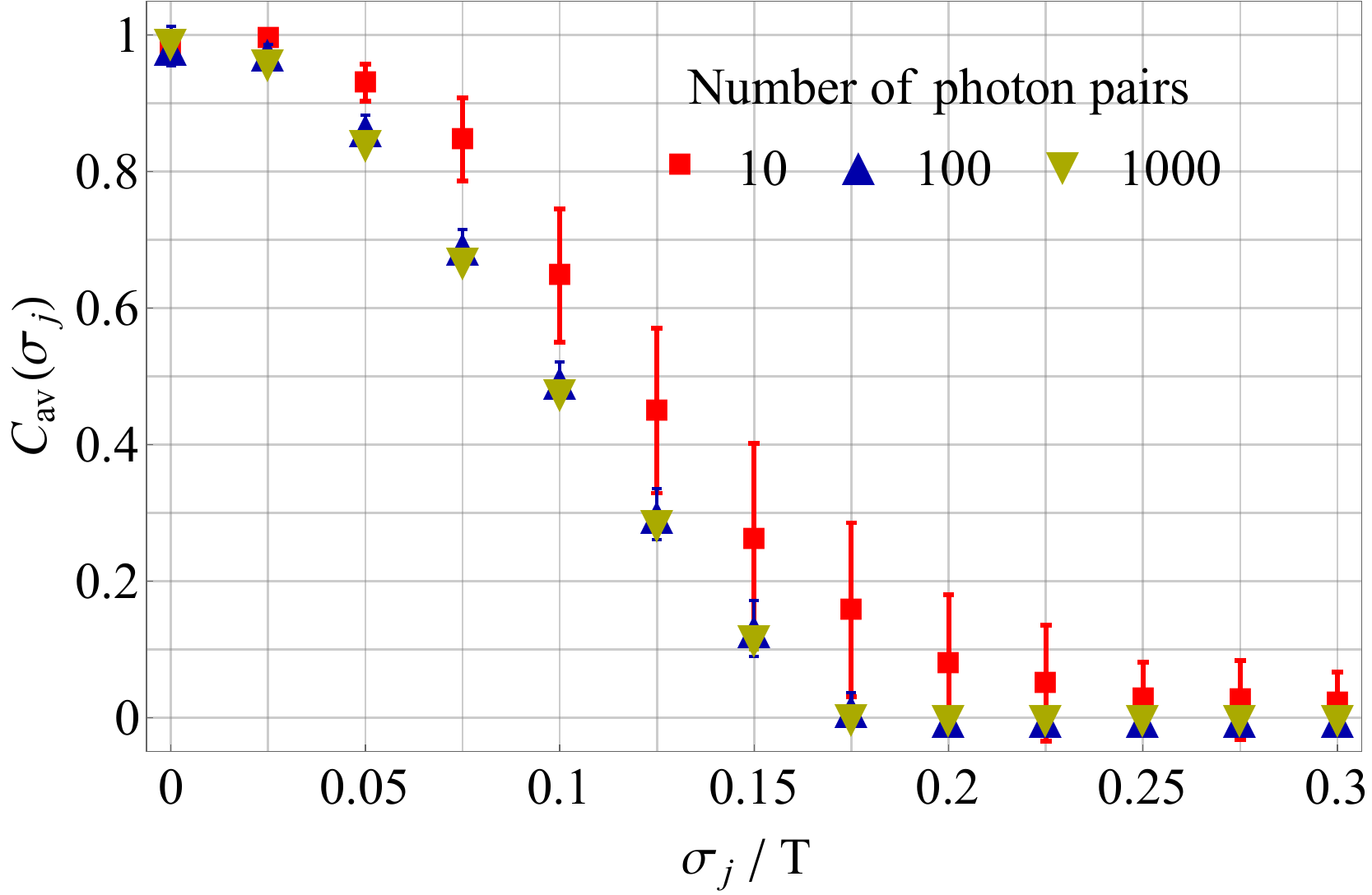}}
	\caption{Plots of $C_{av} (\sigma_j)$ for a sample of $200$ two-qubit entangled states. Three different numbers of photon pairs per measurement were considered.}
	\label{entanglement1}
\end{figure}

We select a sample of $200$ entangled qubits in the form \eqref{3e7} that differ in the value of the relative phase. Each input state is reconstructed based on the measurement scheme described in Sec.~\ref{methequbits}. Then, in order to quantify the amount of entanglement detected by the framework, we compute the average concurrence for the output states. In \figref{entanglement1}, we can observe the plots of $C_{av} (\sigma_j)$ with error bars (one SD) presented as a function of $\sigma_j$ for three numbers of photon pairs generated by the source (per measurement).

First, we can observe the detrimental impact of the detector jitter on entanglement detection. We notice that $C_{av} (0.25\, T) \approx 0$, regardless of the number of photon pairs involved in each measurement. All three plots start at one point, and they converge since for $\sigma_j \geq 0.25\,T$, we obtain zero concurrence. Along the whole interval of $\sigma_j$ the plots corresponding to $\mathcal{N}=100$ and $\mathcal{N}=1\,000$ coincide. However, for $\mathcal{N}=10$, the average concurrence exceeds the other two when $0.025\,T < \sigma_j < 0.225\,T$. In particular, the discrepancy between the plots is most substantial for the middle values of the analyzed interval, i.e., for $0.1\,T \leq \sigma_j \leq 0.175 \,T$ the difference in the average concurrence between $\mathcal{N}=10$ and the other two scenarios is more than $0.15$. The simulations have been repeated several times, and very similar results were obtained. It should be stressed that the concurrence for the states obtained from the low-photon statistics features a great amount of variance, which is presented by error bars. Thus, the results are scattered along a wide range. This implies that with $\mathcal{N}=10$, we cannot guarantee a high-quality entanglement detection since, for a given state, the particular outcome may vary significantly.

\begin{figure}[h]
	\centering
         \centered{\includegraphics[width=0.75\columnwidth]{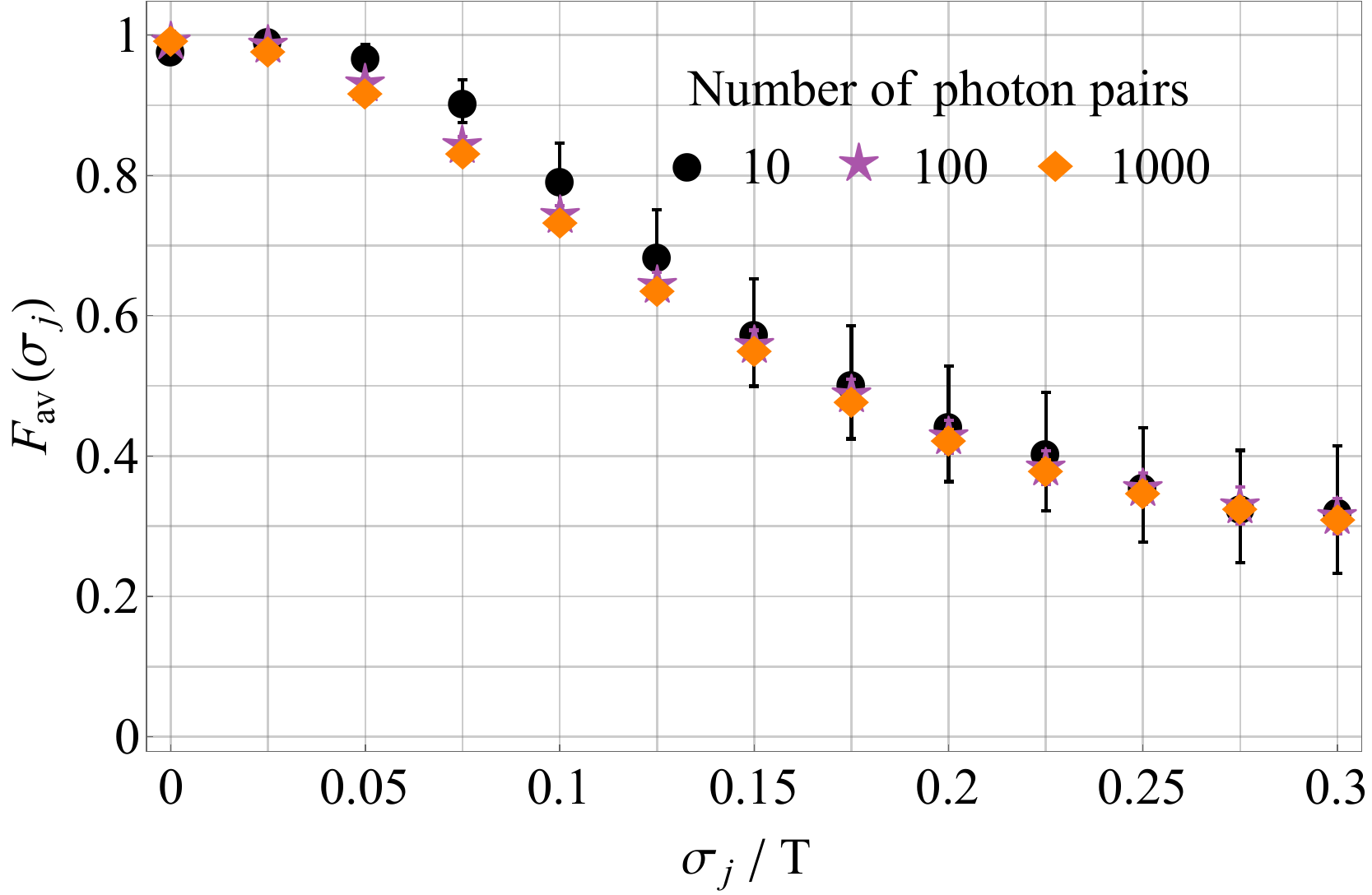}}
	\caption{Plots present the average fidelity, $\mathcal{F}_{av} (\sigma_j)$, in QST of entangled qubits for a sample of $200$ input states of the form \eqref{3e7}.}
	\label{entanglement2}
\end{figure}

For practical implementations, we are usually interested in detecting such amount of entanglement that is sufficient to announce the violation of the Bell-CHSH inequality \cite{Bell1964,Clauser1969}. As far as the concurrence is concerned, one can guarantee the detection of non-classical correlations in the system described by a density matrix $\rho$ if $C[\rho] > 1/\sqrt{2}$ \cite{Verstraete2002,Hu2012}. Based on this condition, we can evaluate the efficiency of the framework in entanglement detection. We know that $99.74\%$ of the values lie within three SDs of the mean, which implies that we need to extend the error bars to guarantee that almost all states satisfy the condition for the Bell-CHSH inequality violation. Then, for $\sigma_j/T = 0.065$, we obtain the following intervals: $C[\rho_{10}] = 0.85\pm0.14$, $C[\rho_{100}] = 0.77\pm0.06$, and $C[\rho_{1000}] = 0.74\pm0.02$, where $\rho_{\mathcal{N}}$ denotes a density matrix obtained from $\mathcal{N}$ photon pairs per measurement. Thus, for all $0 \leq \sigma_j/T \leq 0.065$, we can guarantee entanglement detection with any number of photon pairs considered in this model. As one can notice, the three-sigma interval for $\mathcal{N}=1\,000$ is relatively narrow, which allows one to accurately predict the concurrence, whereas the low-photon scenario leads to more significant statistical dispersion. For $\sigma_j/T = 0.07$, it was checked that within three SDs of the average concurrence, we drop below the threshold, irrespective of the number of photon pairs involved in a single measurement. Therefore, $[0, 0.065\,T]$ can be considered an admissible noise interval of the detector jitter, $\sigma_j$, for entanglement detection.

To further investigate the performance of the QST framework, we computed the average fidelity, which is presented with error bars in \figref{entanglement2}. The plots indicate a modest difference in the average fidelity in favor of the low-photon statistics. However, as we increase $\sigma_j$, we observe significant growth of the SD corresponding to $\mathcal{N}=10$. This implies that state estimation with a low number of photon pairs involves more variability.

\section{Discussion and summary}\label{discussion}

In the article, we introduced a QST framework based on time-continuous measurements generated by unitary dynamics. We considered a realistic scenario with measurement results distorted by both time uncertainty and the Poisson noise. The framework was tested numerically on qubits and entangled qubits.

As for qubits, it was demonstrated that the framework could perform efficiently for a sample of states belonging to the Bloch ball. However, the tomographic scheme is less accurate if we consider only pure states. Since the measurement points collapse inward the Bloch ball, the average fidelity for pure states tomography declines rapidly. In addition, by comparing the results for different numbers of photons per measurement, we discovered that if we utilize $10$ photons, the fidelity in a sample of quantum states features a large variance.

Furthermore, we studied how well two orthogonal states can be distinguished with the framework. Here we identified a deceptive effect associated with a low number of photons per measurement. Due to the Poisson noise and time uncertainty, the plot of the average fidelity corresponding to the single-photon scenario does not converge, which might be incorrectly interpreted as better performance.

In the case of entangled qubits, we studied the precision of state reconstruction quantified by the average fidelity. Furthermore, the amount of entanglement detected by the framework was expressed by the average concurrence. With respect to both aspects, it was discovered that the single-photon scenario provides higher figures, but at the same time, the results feature more variance. Numerical simulations were re-evaluated several times, but the results turned out to be repeatable.

For entangled photon pairs, it was evident that the figures of merit decrease rapidly as we add more time uncertainty. Nonetheless, we identified a noise interval that is admissible for the violation of the Bell-CHSH inequality. If the boundary value of the detector jitter is not exceeded, we can guarantee, within the three-sigma accuracy, the detection of nonlocal correlations in spite of the Poisson noise and time uncertainty.

Last but not least, the framework should be discussed with reference to current technological capabilities. The accuracy of the measurements in the time domain is strictly connected with the quality of single-photon detectors. In our approach, we quantified the temporal uncertainty of the detector by the timing jitter, which was expressed as a relative ratio, compared to the period characterizing the dynamics. In practice, one can apply superconducting nanowire single-photon detectors (SNSPDs) for which the jitter is about $25$ ps \cite{Shcheslavskiy2016,Divochiy2018}. Such off-the-shelf instruments may not be sufficient, but the temporal resolution of photon detectors has constantly been improving, and for state-of-the-art equipment, the jitter can drop down below $3$ ps \cite{Korzh2020}. Thus, when thinking of practical applications of the framework, it should be tested how an available detector will influence state reconstruction for a given polarization evolution that can be generated and controlled in a laboratory.

To conclude, we can state that the framework presented in the article is in accordance with current technological capabilities because both continuous polarization controllers and time-resolved photon detectors are available on the market. In the future, the model can be extended by incorporating the effects caused by dispersion in the fiber. Furthermore, different quantum state estimation techniques can be applied and compared in terms of their efficiency.

\backmatter

\bmhead{Acknowledgments}
I acknowledge K. Sedziak-Kacprowicz and P. Kolenderski, who collaborated with me on prior projects related to time-bin quantum states Ref.~\cite{SedziakKacprowicz2020,Czerwinski2021a}. The present work was inspired by the aforementioned papers. 

\bmhead{Funding}
This research received no external funding.

\section*{Declarations}

\bmhead{Conflict of interest} The author declares that there is no conflict of interest.
\bmhead{Ethics approval} Not applicable
\bmhead{Data availability} Data underlying the results presented in this paper is not publicly available at this time but may be obtained from the author upon reasonable request.
\bmhead{Code availability} Code for the 
numerical simulations described in this paper was written in Mathematica and may be obtained from the author upon reasonable request.

\end{document}